\documentclass[aps,pra,twocolumn,groupedaddress,showpacs]{revtex4}

\usepackage{graphicx}
\usepackage{dcolumn}
\usepackage{bm}
\usepackage{verbatim}
\usepackage{amsmath,amssymb}
\usepackage{relsize,exscale}
\usepackage{color}
\usepackage{hyperref}
\begin{document}

\title{Orbits of hybrid systems as qualitative indicators of quantum dynamics}

\author{N. Buri\'c}
\email[]{buric@ipb.ac.rs}
\author{D. B. Popovi\' c}
\author{M. Radonji\'c}
\author{S. Prvanovi\'c}
\affiliation{Institute of Physics, University of Belgrade,
Pregrevica 118, 11080 Belgrade, Serbia}

\begin{abstract}
Hamiltonian theory of hybrid quantum-classical systems is used to study dynamics of the classical subsystem coupled to
 different types of quantum systems.
It is shown that the qualitative properties of orbits of the classical subsystem
 clearly indicate if the quantum subsystem does or does not have additional conserved observables.
 .

\end{abstract}


\maketitle

\section{Introduction}

Linear Schr\" odinger equation of any quantum mechanical system is equivalent to an integrable Hamiltonian dynamical system
\cite{Arnold,Ham1,Ham2,Ham3,Ham4,Ham5}. As such, the linear Schr\"odinger equation of a bounded system has only periodic or quasi-periodic orbits.
However, integrable systems are exceptional \cite{Ham_book}. Typical Hamiltonian system has also plenty of irregular, i.e. chaotic orbits
 \cite{Ham_book}, but these do not appear in standard quantum mechanics.
 Integrability, or the lack of it, of Hamiltonian dynamical systems is related to the symmetries of the model and to the existence of a sufficient number of integrals of motion.
 The difference between integrable and non-integrable systems is clearly manifested in the qualitative properties of orbits. The former have only
  regular, periodic or quasi-periodic orbits, and in the latter the chaotic orbits dominate.
 Classification of
 quantum system into regular or irregular such as ergodic or chaotic, is possible using different plausible and variously motivated criteria
  without reference to the orbital properties.
 Usually, the criteria are formulated in terms of the properties of the energy spectrum, and the connection with the classical, well developed,
  notions of regular or chaotic dynamics, formulated in terms of orbital properties, is obscured.

    The purpose of our work was  to investigate qualitative properties of orbits of a hybrid quantum-classical system, where the classical part
     is integrable when isolated and the quantum part is  characterized as symmetric or non-symmetric by the existence of constant observables.
      In particular, we want to see if the symmetry, or the lack of it, might be displayed in the qualitative properties of orbits of the classical part. To this end we utilized recently developed Hamiltonian hybrid theory
  of quantum-classical (QC) systems \cite{Elze1,Elze2,Wu, us_PRA3,us_PRA4}. Our main result is that indeed quantum systems, characterized as nonsymmetric
   imply chaotic orbits of the classical degrees of freedom (CDF) coupled to the quantum system. On the other
    hand, CDF show regular dynamics if coupled to a symmetric quantum system, i.e. a quantum system with sufficient number of constant observables.

 One of the first to
  introduce some sort of dynamical distinction between quantum systems was von Neumann \cite{Neumann} with his definition of quantum ergodicity  based on the properties of the Hamiltonian eigenspectrum. Further developments and different approaches to the problems of quantum irregular dynamics can be divided into three groups. The  literature on the topic is enormous, and we shall
    give only a few examples or a relevant review for each of the approaches.
    The most popular was the type of studies
 analyzing  the spectral properties of quantum systems obtained by quantization of chaotic classical systems (see the reviews collected  in \cite{chaology}). Still in the framework
  of systems whose classical analog is chaotic, there were studies of semi-classical dynamics \cite{chaology} and phase space distributions \cite{chaology}. The second group of studies consists of those works where an intrinsic definition of  quantum
   chaoticity is attempted \cite{quantum_erg}. Neither the works in the first nor those in the second group rely on the topological properties of
  pure state orbits of quantum systems. The third group originates from the studies of open quantum systems, and here the properties of orbits
     of an open quantum system are important. Classical property of chaoticity defined in terms of orbital properties was analyzed in quantum systems
      interacting with different types of environments \cite{open1,open2,open3}. It was observed that orbits of such open quantum systems
   in the macro-limit might be chaotic.

    In the next section we shall briefly recapitulate the Hamiltonian theory of hybrid systems. In section 3 we present the hybrid models consisting of qualitatively different pairs of qubits as the quantum part and
    the linear
     oscillator as the classical part.
     Section 4 will describe numerical computations of hybrid dynamics and our main results. Brief summary  will be given in section 5.


\section{Hamiltonian hybrid theory}

There is no unique generally accepted theory of interaction between micro and macro degrees of freedom, where
the former are described by quantum and the latter by classical theory (see \cite{Elze1} for an informative
review).
 Some of the suggested hybrid theories are mathematically
inconsistent, and ``no go" type theorems have been formulated \cite{Salcedo}, suggesting that no consistent
hybrid theory can be formulated. Nevertheless, mathematically consistent but inequivalent hybrid theories exist
\cite{Elze1,Diosi,Royal,Hall, usPRA5}.

The Hamiltonian hybrid theory, as formulated and discussed for example in \cite{Elze1,us_PRA3,us_PRA4},
has many of the properties commonly expected of a good hybrid theory, but has also some controversial features. It's physical content is equivalent to the standard mean field approximation, but it is formulated entirely in terms of the Hamiltonian framework, which provides useful insights such as the one presented in this communication.
The theory is based on the equivalence of the Schr\" odinger equation on ${\cal H}^N$ and the corresponding Hamiltonian system on ${\mathbb R}^{2N}$.
The Riemannian $g$ and the symplectic $\omega$  structures on the phase space ${\cal M}_q={\mathbb R}^{2N}$ are given by the real and imaginary parts
of the Hermitian scalar product on ${\cal H}^N$: $\langle \psi|\phi\rangle=g(\psi,\phi)+i\omega(\psi,\phi)$. Schr\" odinger equation in an abstract
 basis $\{|n\rangle\}$ of ${\cal H}^N$
 \begin{equation}\label{sch}
i\hbar \frac{\partial c_n}{\partial t}=\sum_ m H_{nm}c_m
\end{equation}
where $|\psi\rangle=\sum_n c_n |n\rangle$ and $H_{nm}=\langle n|\hat H|m\rangle$ is equivalent to Hamiltonian equations
\begin{equation}\label{hameq}
\dot x_n=\frac{\partial H(x,y)}{\partial y_n},\quad \dot y_n=-\frac{\partial H(x,y)}{\partial x_n}
\end{equation}
where $c_n=(x_n+iy_n)/\sqrt{2\hbar}$ and
\begin{equation}\label{hamf}
H(x,y)=\langle \psi_{xy}|\hat H|\psi_{xy}\rangle,
\end{equation}
where $(x,y)$ stands for $(x_1,x_2\dots x_N,y_1,y_2\dots y_N)$.
Only quadratic functions $A(x,y)$ of the form $A(x,y)=\langle \psi_{xy}|\hat A|\psi_{xy}\rangle$ are related to the physical observables $\hat A$.
 In particular, the canonical coordinates $(x,y)$ of quantum degrees of freedom (QDF) do not have such interpretation.

 Hamiltonian hybrid theory uses the Hamiltonian formulations of quantum and classical dynamics, and couples the classical and quantum systems as they would be coupled in the theory of Hamiltonian systems. The phase space of QC system is given by the Cartesian product
 \begin{equation}\label{phasespace}
 {\cal M}_{qc}={\cal M}_q\times {\cal M}_c,
 \end{equation}
and the total Hamiltonian is of the form
\begin{equation}\label{htot}
H_{qc}(x,y,q,p)=H_q(x,y)+H_{cl}(q,p)+H_{int}(x,y,q,p).
\end{equation}
The dynamical equations of the hybrid theory are just the Hamiltonian equations with the Hamiltonian (\ref{htot}).

Observe two fundamental properties of the Hamiltonian hybrid theory: a) There is no entanglement between QDF and CDF and b) the canonical coordinates
 of CDF have the interpretation of conjugate physical variables and have sharp values in any pure state $(x,y,q,p)$ of the hybrid. Hamiltonian theory of hybrid systems can be developed starting from the Hamiltonian formulation of a composite
quantum system and imposing a constraint that one of the components is behaving as a classical system
\cite{us_PRA3}.

\section{Qualitatively different quantum systems coupled to the classical harmonic oscillator}

 We shall consider the following three examples of quantum system with different symmetry properties. All three examples involve a pair of interacting qubits, where $\sigma_{x,y,z}^{1,2}$ denote $x,y$ or $z$ Pauli matrix of the qubit 1 or the qubit 2, and $\omega,\>\mu$ and $\beta$ are parameters. The simplest is given by
  \begin{equation}\label{hs}
    \hat H_s=\hbar\omega  \sigma_z^1+\hbar\omega \sigma_z^2+\hbar\mu \sigma_z^1 \sigma_z^2.
    \end{equation}
The system has two additional independent constant observables $\sigma_z^1$ and $\sigma_z^2$ corresponding to the $SO(2)\times SO(2)$ symmetry
 of the model. Next two models are examples of non-symmetric systems. The system
  \begin{equation}\label{hns1}
   \hat H_{ns1}=\hat H_s+\hbar\beta \sigma_y^1
    \end{equation}
    has only $\sigma_z^2$ as the additional constant observable, and in the system
   \begin{equation}\label{hns2}
    \hat H_{ns2}=\hbar\omega  \sigma_z^1+\hbar\omega \sigma_z^2+\hbar\mu \sigma_x^1 \sigma_x^2,
    \end{equation}
    there are no additional dynamical constant observables.
 Let us stress that the Hamiltonian systems with the Hamiltonian functions given by $\langle\psi|\hat H|\psi\rangle$ are integrable with only the
  regular (non-chaotic) orbits irrespective of their symmetry properties.


  \begin{figure*}[t]
\centering
\includegraphics[scale=0.33]{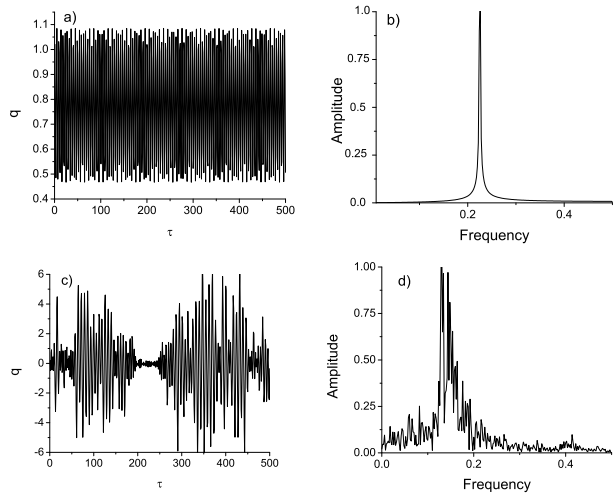}
\caption{Figures illustrate the time series $q(\tau)$ (a,c) and the corresponding amplitudes of the Fourier spectra (b,d), of the classical oscillator subpart of the hybrid system with the quantum subpart given by symmetric (\ref{hsf}) (a,b) and non-symmetric (\ref{hns2f}) (c,d) systems. The values of the parameters are
$\omega=1,\> \mu=5, m=k=1,\> c_1=15,c_2=1$.
\label{Fig1}}
\vspace{-0.6cm}
\end{figure*}

\begin{figure*}[t]
\centering
\includegraphics[scale=0.33]{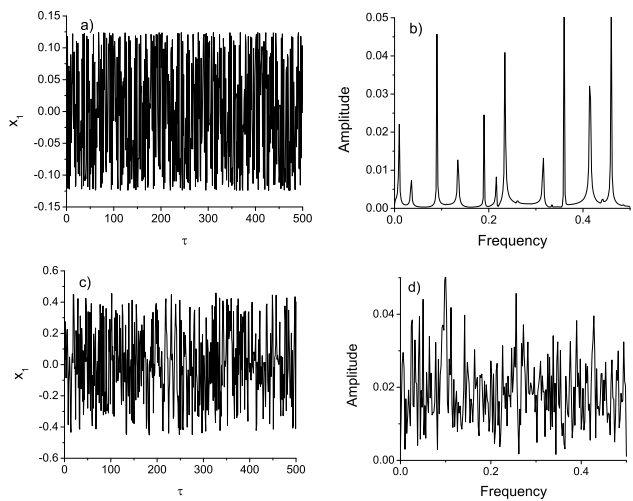}
\caption{ Figures illustrate the time series (a,c) and the corresponding amplitudes of the Fourier spectra (b,d), of the $x_1$ canonical coordinate of the quantum subpart of the hybrid system  given by  symmetric $H_{s}$(\ref{hsf}) (a,b) and non-symmetric $H_{ns2}$(\ref{hns2f}) (c,d) systems. The values of the parameters
are the same as in fig.1.
\label{Fig2}}
\vspace{-0.6cm}
\end{figure*}
\begin{figure*}[t]
\centering
\includegraphics[scale=0.33]{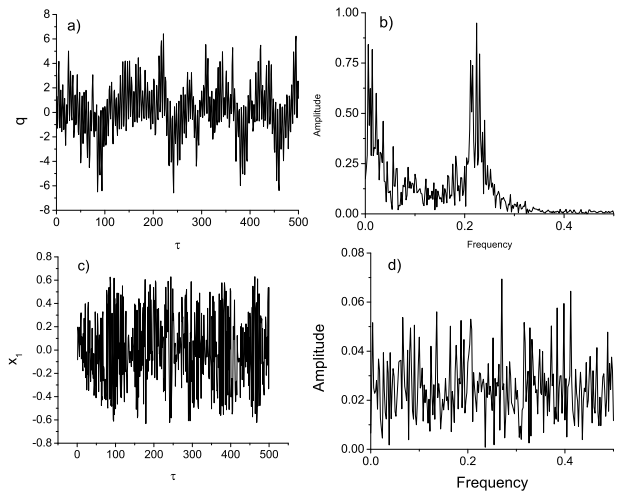}
\caption{ Figures illustrate the time series $q(\tau)$ (a) and $x_1(\tau)$ and the corresponding amplitudes of the Fourier spectra (b,d). The
 hamiltonian is non-symmetric $H_{ns1}$(\ref{hns1f}). The values of the parameters are the same as in fig.1.
\label{Fig3}}
\vspace{-0.6cm}
\end{figure*}

The Hamilton functions corresponding to the three quantum systems (\ref{hs}),(\ref{hns1}) and (\ref{hns2}) are given by  the general rule (\ref{hamf}). In the computational basis
 $|1\rangle=|1,1\rangle,\> |2\rangle=|1,-1\rangle, \> |3\rangle=|-1,1\rangle, \> |4\rangle=|-1,-1\rangle$, where
  for example $|1,1\rangle=|1\rangle\otimes|1\rangle$ and $|\pm 1\rangle$ are the eigenvectors of $\sigma_z$,
  the Hamilton functions are
   \begin{eqnarray}\label{hsf}
 && H_{s}(x,y)=\omega (x_1^2+y_1^2-x_4^2-y_4^2)\nonumber\\&+&\frac{\mu}{2} (x_1^2-x_2^2-x_3^2+x_4^2+y_1^2-y_2^2-y_3^2+y_4^2),
  \end{eqnarray}
  \begin{eqnarray}\label{hns1f}
  &&H_{ns1}(x,y)=\omega (x_1^2+y_1^2-x_4^2-y_4^2)\nonumber\\
  &+&\frac{\mu}{2} (x_1^2-x_2^2-x_3^2+x_4^2+y_1^2-y_2^2-y_3^2+y_4^2)\nonumber\\
  &+&\beta (y_3 x_1 + y_4 x_2 - y_1 x_3 - y_2 x_4)
  \end{eqnarray}
  and
 \begin{eqnarray}\label{hns2f}
  &&H_{ns2}(x,y)=\omega (x_1^2+y_1^2-x_4^2-y_4^2)\nonumber\\
  &+&\mu (x_2x_3+x_1x_4+y_2y_3+y_1y_4).
  \end{eqnarray}
  Observe that, due to the $1/\sqrt{2\hbar}$ scaling of the canonical coordinates $(x,y)$, $\hbar$ does not appear in the Hamilton's functions
  (\ref{hsf}), (\ref{hns1f}) and (\ref{hns2f}) nor in the corresponding Hmilton's equations and their solutions $x(t)\dots$. Of course, $\hbar$ reappears in the functions $\langle\sigma_{x}^1\rangle\dots$.

The classical system that we want to couple with quantum systems (\ref{hsf}), (\ref{hns1f}) or (\ref{hns2f}) is one-dimensional linear oscillator with the Hamiltonian
\begin{equation}\label{hosc}
H_{cl}(q,p)=\frac{p^2}{2m}+kq^2,
\end{equation}
which of course has only regular periodic orbits.

The $QC$ interaction term is taken to be such that it does not interfere with the existence of operators commuting with the Hamiltonian of the quantum  part. In other words, the operator $\hat H_q+\hat H_{int}$  has the same additional constant observables as the quantum part $\hat H_q$.
Furthermore,  $\hat H_{int}$ must depend on observables of the qubit 1 and of the qubit 2.
  For example
$\hat H_{int}= q(c_1\hbar\sigma_z^1+c_2\hbar\sigma_z^2)$ implying  $H_{int}(x,y,q,p)=q(c_1\hbar\langle\sigma_z^1\rangle+c_2\hbar\langle\sigma_z^2\rangle)$ or explicitly
\begin{eqnarray}\label{hintf}
&&H_{int}=\frac{c_1 q}{2} (x_1^2+x_2^2-x_3^2-x_4^2+y_1^2+y_2^2-y_3^2-y_4^2)\nonumber\\
&+&\frac{c_2q}{2} (x_1^2-x_2^2+x_3^2-x_4^2+y_1^2-y_2^2+y_3^2-y_4^2).
\end{eqnarray}
The total Hamiltonian is given by the sum of (\ref{hosc}), (\ref{hintf}) and one of (\ref{hsf}), (\ref{hns1f}) or (\ref{hns2f}). Observe that the functions $\langle \sigma_z^1\rangle$ and $\langle \sigma_z^2\rangle$ are constants of motion for the hybrid
$H_s+H_{int}+H_{cl}$, as is the function $\langle \sigma_z^2\rangle$ constant for the hybrid $H_{ns1}+H_{int}+H_{cl}$. Thus, $H_{int}$ given by (\ref{hintf}) satisfies
 the general condition that we impose on the QC interaction.


\section{Numerical computations and the results}

Hamiltonian equations are solved numerically and the dynamics of CDF, illustrated in fig. 1 and fig. 3a,b and of QDF illustrated  in fig. 2 and fig. 3c,d, is observed in the
 cases corresponding to the symmetric or non-symmetric  quantum parts for different values of the parameters $\mu$ and $c$.
 Let us first stress again that if there is no classical system then all orbits are  regular for either of the quantum systems.
 On the other hand the hybrid system displays different behavior.
Consider first the time series generated by the CDF.  Figures 1a,b,c,d and figures 3a,b show the time series $q(\tau)$ (fig. 1a,c and fig. 3a), where $\tau=\omega t$ is the dimensionless time,
 and the corresponding Fourier amplitude spectra (fig. 1b,d and fig.3b). Fig. 1a,b are obtained with the quantum symmetric system  (\ref{hsf}), fig.\ 1c,d with quantum non-symmetric system (\ref{hns2f}) and fig. 3 with quantum non-symmetric system  (\ref{hns1f}).
  Obviously, the orbits of the CDF are periodic, with single frequency, in the symmetric case, and chaotic with a broad-band spectrum
  in the non-symmetric cases.  We can conclude that the qualitative properties of orbits of a classical system coupled with a quantum system are
  excellent indicators of the symmetries of the quantum system.


Consider now the dynamics of QDF illustrated in fig. 2a,b,c,d. and fig. 3c,d by plotting the time series generated by $x_1(t)$ and the corresponding Fourier amplitudes spectra. Qualitatively the same properties are displayed by dynamics of other canonical coordinates $x_2,x_3,x_4,y_1,y_2,y_3,y_4$ or, for example, by the dynamics of expectation values $\langle \sigma_x^1(t)\rangle,\dots$. Again, the time series are regular if the quantum systems are
 symmetric and are chaotic in the quantum non-symmetric case. The same conclusion is obtained with $H_{ns2}$ replaced by $H_{ns1}$.
  We can conclude that the orbits of the hybrid system, are regular or chaotic, in the sense of Hamiltonian dynamics, depending on the quantum subpart being symmetric or non-symmetric. Thus, the relation between symmetry and existence of independent constants of motion on one hand and the
   qualitative properties of orbits on the other, which is the characteristic feature of classical mechanics and is not a feature of
    isolated quantum systems, is restored by appropriate coupling of the quantum and a classical integrable system.

  Observe that such behavior can not be obtained by coupling two quantum systems (instead of quantum-classical coupling). In this case, and
   even for the simplest quantum system in place of the classical one,
   the phase space of the quantum composite system is much larger than ${\cal M}_{qc}$ because of the degrees of freedom corresponding to the possibility of entanglement, and the total system is always linear. All degrees of freedom of a quantum-quantum system in the Hamiltonian formulation display only regular dynamics, independently of the symmetries of the quantum Hamiltonian. On the other hand, the hybrid systems are
    nonlinear, due to the QC coupling and the phase space of the form (\ref{phasespace}), and the relation between the symmetries and the qualitative properties of orbits is like in the general Hamiltonian theory.

  Explanation of the observed properties relies on the fact that the five degrees of freedom hybrid Hamiltonian system with quantum symmetric subpart has enough independent constants
   of motion in involution. These are  given by $H(x,y,q,p),H_{s}(x,y),\langle\sigma_z^1\rangle,\langle\sigma_z^2\rangle$ and the norm of the state of the quantum subpart. On the other hand $H_{ns1}+H_{int}+H_{cl}$, or $H_{ns2}+H_{int}+H_{cl}$ do not have enough such constants of motion since the
   quantum part $\hat H_{ns2}$ does not commute with $\sigma_z^1$ and $\sigma_z^2$ and $\hat H_{ns1}$ with $\sigma_z^1$. Only $H_s+H_{int}+H_{cl}$ is
    integrable while those obtained with non-symmetric quantum  subparts are not and thus have some chaotic orbits.


\section{Summary}

In summary, we have shown that the orbits of an integrable classical system when coupled to a quantum system in an appropriate way remain regular or become chaotic depending on the presence or lack of symmetries in the quantum part. To this end we used the Hamiltonian theory of quantum-classical systems and examples of qubit systems. The first fact is
  an important restriction on our work. On the second point, the nature of our results is qualitative and is therefore expected to be valid generically, and not only for the considered examples. Considering the choice of Hamiltonian theory to describe QC interaction, we were motivated by the mathematical consistency of the theory and the fact that the theory describes orbits of pure states of a deterministic Hamiltonian system. There are other consistent hybrid theories, but they are either formulated in terms of probability densities \cite{Royal, Hall} or in terms of
   stochastic pure state evolution \cite{Diosi,usPRA5}. Of course, the significance of our result could be properly judged only after the status of
    Hamiltonian hybrid theory is sufficiently understood.

\begin{acknowledgments}
We acknowledge support of the Ministry of  Education and Science of the Republic of Serbia, contracts
No. 171006, 171017, 171020, 171038 and 45016 and COST (Action MP1006).
\end{acknowledgments}

\end{document}